\documentclass[a4paper,11pt]{article}

\usepackage{multicol}
\usepackage{graphicx}
\usepackage{setspace}
\usepackage[
   pdftitle={Inelastic electron scattering by metastable hydrogen atoms in the presence of a laser field}, 
   pdfauthor={Author One, Author Two, and so on},         
   pdfkeywords={photons, electrons, atoms, collisions},   
   colorlinks,
   linkcolor=blue,
   urlcolor=blue,
   citecolor=blue,
   a4paper=true,
   ]{hyperref}

\newcommand{\abstracttitle}[1]{
 \begin{center}{\Large {\bf #1}}\end{center}
}
\newcommand{\authors}[1]{
 \vspace*{-0.3cm}
 \begin{center} {\bf #1} \end{center}
 \vspace*{-0.3cm}
}
\newcommand{\addresses}[1]{
 \begin{center} {\small #1} \end{center}
}
\newcommand{\synopsis}[1]{
 \begin{center}
 \setstretch{0.75}
 \begin{minipage}[t]{16cm}
   {\footnotesize {\bf Synopsis} #1 }
 \end{minipage}
 \setstretch{1.0}
 \end{center}
}
\newcommand{\abstracttext}[1]{
 \vspace*{-0.3cm}
 \columnsep0.75cm
 \begin{multicols}{2} #1 \end{multicols}
}

\newcommand{\picturelandscape}[2]{
 \vspace*{0.5cm}
 \centerline{
  \includegraphics*[width=7.8cm,angle=#1]{#2}
 }
}
\newcommand{\capt}[2]{
 \vspace*{-0.3cm}
 \begin{center}
 \begin{minipage}[t]{7.8cm} {\small {\bf Figure~#1}.~#2} \end{minipage}
 \end{center}
 \vspace*{0.3cm}
}

\newcommand{\writeto}[1]{
 \hspace*{-2.5mm} \footnote{E-mail: \href{mailto:#1}{#1}}\hspace*{-1.5mm}
}

\begin{document}

\abstracttitle{Laser-assisted inelastic electron scattering by excited hydrogen atoms}

\authors{
Gabriela Buica\writeto{buica@spacescience.ro}
}

\addresses{
 Institute of Space Sciences, P.O. Box MG-23, Ro 77125,
Bucharest-M\u{a}gurele, Romania}

\synopsis{
We theoretically study the \textit{inelastic} electron-H($2s$) scattering in a linearly
 polarized laser field in the domain of high scattering energies and
 moderate field intensities.
We investigate the angular distributions and the resonance structure of the  differential
scattering cross sections.
}
\abstracttext{

The purpose of this work is to study the  \textit{inelastic} scattering
of fast electrons  by hydrogen atoms in the metastable $2s$ state in
the presence of a linearly polarized laser field.
For elastic scattering the initial and final states of the target are the same,
 while for inelastic scattering the final state differs from the initial one.
We consider the excitation of metastable hydrogen to an arbitrary state
accompanied by one-photon absorption.
It is  important to evaluate the contribution of the laser-assisted
\textit{inelastic} electron-atom scattering to the total electron energy spectrum since
in experimental studies  it might be quite difficult
 to separate the signal of elastic and inelastic scattering
channels \cite{joachain}.

Since the scattering process under investigation is a three-body problem,
i.e., projectile, atomic target, and photon, the
theoretical treatment becomes very complicated and we make several assumptions.
We focus on moderate field intensities and fast projectile electrons in order to neglect
the exchange scattering and the second Born approximation in the scattering potential.
First, the interaction between the projectile electron and the laser field
is described exactly by a Gordon-Volkov wave function.
Second, the dressing of the hydrogen atom by the laser field is described
within the first-order time-dependent perturbation theory in the
field \cite{vf1}.
Finally, the interaction between the fast projectile electron and the hydrogen
atom is treated in the first Born approximation  \cite{b-j}.
Using the approach described in  \cite{ac} we have obtained an
\textit{analytical formula} for the differential cross section (DCS)
in the laser-assisted inelastic e-H($2s$) scattering.

We analyze the angular distributions and the resonance structure of the
DCS's for the atomic excitation of the $n=4$ subshells.
In Figure~1 we show the DCS's with respect of the photon energy
for one-photon absorption  corresponding to the $2s \to 4s$ (short dashed line),
 $2s \to 4p$ (dot-dashed line),  $2s \to4d$ (long dashed line),
and $2s \to 4f$ (solid line) excitation processes of H$(2s)$, in the scattering
 geometry where the laser field  is linearly polarized in the same direction along
the momentum of the ingoing electron.
The DCS's are normalized to the laser intensity and calculated  at the
incident projectile energy  of $500$ eV
and the scattering angle  $\theta = 5^{\circ}$.

\picturelandscape{0}{fig5.eps}
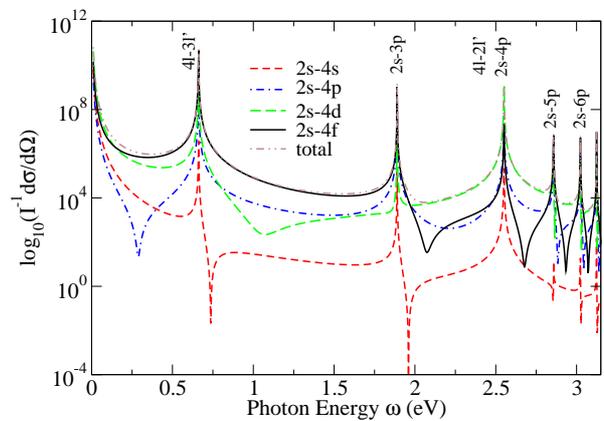
\capt{1}{Differential cross sections for
\textit{inelastic}  laser assisted e($E_i$)+H($2s$)$\to$ e($E_f$)+H($4l$)
 scattering process with the excitation
 of the  $4l$ subshells as a function of the photon energy.}

Our preliminary results indicate that the laser-dressing effects are
 significantly stronger than for the elastic $2s \to 2s$ collision \cite{acgabi3}
 (static dipole polarizability $\alpha_{2s}$=120 a.u),
in particular at small scattering angles.
For the studied range of photon energy our analysis shows important
differences from the case of laser-assisted elastic scattering,
 that are mainly due to the intermediate resonances in
the laser-atom interaction.

}

\end{document}